\documentclass{ws-ijmpa}
\def\be{\begin{equation}}
\def\ee{\end{equation}}
\def\bea{\begin{eqnarray}}
\def\eea{\end{eqnarray}}

\def\p{\partial}

\def\a{\alpha}

\def\b{\beta}

\begin{document}

\markboth{F. C. Mena, R. Tavakol \& M. Bruni} {Second 
Order Perturbations of
Flat Dust FLRW}

\catchline{}{}{}

\title{SECOND ORDER PERTURBATIONS OF FLAT DUST FLRW UNIVERSES WITH A COSMOLOGICAL CONSTANT
}

\author{\footnotesize FILIPE C. MENA\footnote{email: fmena@math.uminho.pt}}

\address{Departamento de Matem\'atica,
Universidade do Minho,\\ Campus de Gualtar, 4710 Braga, Portugal}

\author{REZA TAVAKOL}

\address{Astronomy Unit, School of Mathematical Sciences, Queen Mary,\\
University of London, Mile End Road, London E1 4NS, U.K.}

\author{MARCO BRUNI}

\address{Institute of Cosmology and Gravitation, Mercantile House,\\ 
University of Portsmouth, Portsmouth PO1 2EG, U.K.}
\maketitle

\pub{Received 5 July 2002}{}
%%%%%%%%%%%%%%%%%%%%%%%%%%%%%%%%%%%%%%%%%%%%%%%%%%%%%%%%%%%%
\begin{abstract}
%%%%%%%%%%%%%%%%%%%%%%%%%%%%%%%%%%%%%%%%%%%%%%%%%%%%%%%%%%%%
We summarize recent results concerning the evolution of second
order perturbations in flat dust irrotational FLRW models with
$\Lambda\ne 0$. We show that asymptotically these perturbations
tend to constants in time, in agreement with the cosmic no-hair
conjecture. We solve numerically the second order scalar
perturbation equation, and very briefly discuss its all time
behaviour and some possible implications for the structure
formation.

\keywords{Mathematical cosmology; non-linear perturbations; cosmic
no-hair conjecture; structure formation.}
\end{abstract}
%%%%%%%%%%%%%%%%%%%%%%%%%%%%%%%%%%%%%%%%%%%%
\section{Introduction}
%%%%%%%%%%%%%%%%%%%%%%%%%%%%%%%%%%%%%%%%%%%%%
First order perturbations of
Friedmann-Lema\^{\i}tre-Robertson-Walker (FLRW) models fail to
account for a number important relativistic properties that 
arise when non-linearities are taken into account. An
important example of this is the so called mode coupling. Even if
one considers purely scalar perturbations, a comoving scale $k$ at
second order is sourced by any other scale $k^\prime$. In
addition, the scalar, vector and tensor perturbations which are
decoupled to the first order, become coupled once nonlinear
perturbations are taken into account. Thus for example taking into
account such couplings at second order results in the initial pure
scalar metric perturbations to necessarily generate second order
vector and tensor perturbation modes (see e.g.\cite{MMB}). In
view of this, it is important to develop higher order perturbation
schemes which go beyond the first order and can thus account for
the physical effects which are not taken into account up to the
linear order. This would also improve the level of accuracy of
results as well allowing the study of stability of the results
obtained using first order perturbations.

Among concrete motivations for the use of a second order
perturbative scheme is the need for more accurate results from the
new generation of gravitational wave detectors as well as the
increased computational precision required in order to analyse the
cosmic microwave background anisotropies.\cite{Pyne-Carroll,Mollerach-Matarrese} 
Both these problems are
likely, in future, to require analyses that go beyond the linear
order. In addition, the non-linear analysis in the context of
cosmological modeling is also relevant on scales much smaller than
the cosmological horizon where it might substantially modify the
first order results\cite{Campos-Tomimura} and therefore, as we
shall also see below, become important for the study of the
structure formation. Furthermore, as far as observations are
concerned, second order effects can be important on small scales
in order to account for non-linear effects such as the Rees-Sciama\cite{Rees-Sciama} 
and gravitational lensing effects (see e.g.\cite{Schneider-etal}). 
For a list of references concerning these
questions we refer the reader to\cite{Mollerach-Matarrese}.

Second order perturbations have so far not been widely studied in
general relativity. In the cosmological context they seem to have
been first employed by Tomita\cite{Tomita} to study the evolution
of scalar perturbations in the Einstein--de-Sitter model using a
synchronous gauge. This study was repeated by Mataresse et. al.\cite{Matarrese-silent},
who obtained similar results using a
comoving approach and Russ et al.\cite{Russ-etal} who have
included second order terms resulting from a coupling between
growing and decaying scalar perturbation modes.
%In a number of papers Salopek, Stewart and
%collaborators have developed a gradient expansion
%technique in order to calculate perturbations
%of arbitrary order in FLRW backgrounds.\cite{Salopek,Salopek-Stewart,Salopek-etal}
%but its practical application seems difficult
% (see e.g. Matarrese and Terranova\cite{Matarrese-Terranova}).
There have also been recent works\cite{Mukhanov-etal96} showing
that the back reaction of the second order perturbations (to be
precise, perturbations quadratic in first order terms) are
important in early universe scenarios. Furthermore,
%In fact, the recent work of Bassett and
%collaborators\cite{Bassett2} also shows that in
it has been shown that in inflationary models the magnitude of the
second order perturbations can be comparable to those in the first
order and therefore a non-linear perturbative analysis may be
crucial.\cite{Bassett2}

The domain of applicability of the theory of second order
perturbations that we shall describe is that of small
perturbations about an homogeneous and isotropic background. We
shall employ it as an approximate framework in order to study the
effect of non-linearities in structure formation scenarios as well
as the study of the stability of cosmological results obtained
using first order perturbative schemes with respect to nonlinear
perturbations.
%orts  to the irst  test of cosmological attractors. %, such as the de-Sitter %space-time.

In this paper we shall summarise recent results concerning
%where,
%motivated
%by the cosmic no-hair conjecture \cite{Gibbons-Hawking},
%an analysis was made
the exact asymptotic solutions of the second order perturbation
equations. An immediate consequence of these results was to prove
the non-linear asymptotic stability of the de-Sitter attractor\cite{BMT2001}, 
and thus generalise the cosmic no-hair conjecture.\cite{Gibbons-Hawking} 
In general, however, the perturbation
equations cannot be solved analytically. We shall show preliminary
results concerning the numerical integration of the scalar
perturbation equation for all times. Finally, we discuss possible
implications on the structure formation. We use units in which
$8\pi G=c=1$. Greek indices take values $1,2,3$.
%%%%%%%%%%%%%%%%%%%%%%%%%%%%%%%%%%%%%%%%%%%%%%
\section{Perturbation Equations}
%%%%%%%%%%%%%%%%%%%%%%%%%%%%%%%%%%%%%%%%%%%%%%
We consider a homogeneous and isotropic background space-time
given by a flat FLRW metric with an irrotational dust source
field.
%We can take the $3$-metric defined on spacelike
%hypersurfaces orthogonal to the velocity vector of the flow
%$u_a=\delta^0_{~a}$:
%\begin{equation}
%h_{ab}=g_{ab}+u_a u_b
%\end{equation}
The relevant equations for the expansion $\theta^\a_{~\b}$ in
presence of the cosmological constant $\Lambda$ are given by the
evolution equation, the Raychaudhuri equation
%the momentum constraint
and the energy constraint:
\begin{eqnarray}
&& \dot{\theta}_{\a\b}+\theta\theta_{\a\b}+R^*_{\a\b}=
\frac{1}{2}\rho\delta_{\a\b}+\Lambda\delta_{\a\b} \label{evolutiona}\\
&& \dot{\theta}+\theta^{\a\b}\theta_{\a\b}+\frac{1}{2}\rho=\Lambda \label{Ray}\\
%&&  \theta^\a_{~\b;\a}=\theta_{,\b}\\
&&\theta^2-\theta^{\a\b}\theta_{\a\b}+R^*=2(\rho+\Lambda)
\label{energy}
\end{eqnarray}
%\begin{equation}
%\label{evolutiona}
%\dot{\theta}_{\a\b}+\theta\theta_{\a\b}+R^*_{\a\b}=
%\frac{1}{2}\rho\delta_{\a\b}+\Lambda\delta_{\a\b},
%\end{equation}
%the Raychaudhuri equation
%\begin{equation}
%\label{Ray}
%\dot{\theta}+\theta^{\a\b}\theta_{\a\b}+\frac{1}{2}\rho=\Lambda,
%\end{equation}
%the momentum constraint
%\be
%\label{mra1}
%\theta^\a_{~\b;\a}=\theta_{,\b}
%\ee
%and the
%energy constraint equation.
%\begin{equation}
%\label{energy}
%\theta^2-\theta^{\a\b}\theta_{\a\b}+R^*=2(\rho+\Lambda),
%\end{equation}
where the dot denotes a derivative with respect to coordinate time
$t$, $\rho$ is the matter density and $R^*$ is the 3-Ricci scalar.
%and we recall that
%$\theta^\a_\b=u^\a_{;\b}=\frac{1}{2}h^{\a\gamma}\dot{h}_{\gamma\b}$.
We use the notation of Matarrese, Mollerach and Bruni\cite{MMB}
and generalise their formalism to the case with nonzero $\Lambda$.

Using a conformal rescaling we can write the metric as
\[
ds^2=a^2(\tau)(-d\tau^2+\gamma_{\a\b}dx^\a dx^\b),
\]
where $a$ is the scale factor. We work in the synchronous gauge
(which for dust is also comoving) in which the first order metric
perturbations can be written as
\begin{equation}
\label{mare1}
\gamma_{\alpha\beta}=\delta_{\alpha\beta}+\gamma^{(1)}_{\alpha\beta}
%\end{equation}
~~~~~with~~~~~
%\begin{equation}
\label{mare2}
\gamma^{(1)}_{\alpha\beta}=-2\phi\delta_{\alpha\beta}+D_{\a\b}\chi+\pi_{\alpha\beta},
\end{equation}
where $D_{\a\b}=\p_\a \p_\b-\frac{1}{3}\delta_{\a\b}\nabla^2$, the
superscript $^{(1)}$ denotes first order quantities, $\phi$ and
$\chi$ are the trace and tracefree parts of the scalar
perturbation modes and the tensor modes are represented by
$\pi_{\a\b}$, which is transverse and divergence free. In the
linear theory, the scalar, vector and tensor modes are independent
and for an irrotational space-time one can set the vector modes
equal to zero.

We recall that the evolution equation for the tensor modes can be
obtained from the tracefree part of (\ref{evolutiona}) in the form
\begin{equation}
\label{evolution-tensor}
\pi''_{\alpha\beta}+2\frac{a'}{a}\pi'_{\alpha\beta}-\nabla^2\pi_{\alpha\beta}=0,
\end{equation}
where the prime denotes a derivative with respect to conformal
time $\tau$. Using tensor harmonics, which are eigenfunctions of
the
%covariantly defined
Laplace operator, one can set $\nabla^2\pi_{\a\b}=-k^2\pi_{\a\b}$,
where $k$ is the order of the harmonic.

The first order scalar perturbation equation can be obtained from
the continuity equation (which in our case reads
$\dot\rho+\theta\rho=0$), together with (\ref{evolutiona}) and
(\ref{energy}), to give
\begin{equation}
\label{evolution-delta} \delta''+\frac{a'}{a}\delta'-\frac{1}{2}
a^2\rho_b\delta=0,
\end{equation}
where $\delta=(\rho-\rho_b)/\rho_b$ and $\rho_b$ denotes the
background matter density. One can further make use of the freedom
of the synchronous coordinate system to specify
%$\nabla^2\chi_0+2(\rho_0-\rho_b)=0$,
%where the subindex $"0"$ denotes evaluation at an initial time $\tau=\tau_0$.
%which makes
$\delta=-\frac{1}{2}\nabla^2\chi$. This then completely fixes the
gauge.

Having recalled the first order perturbation equations we shall
now consider the second order case. The conformal spatial metric
tensor of the second order perturbed space-time can be written, as
\be \label{mare31}
\gamma_{\alpha\beta}=\delta_{\alpha\beta}+\gamma^{(1)}_{\alpha\beta}+
\frac{1}{2}\gamma^{(2)}_{\a\b}
%\ee
~~~~~with~~~~~
%\be
\label{mare32}
\gamma^{(2)}_{\a\b}=-2\phi^{(2)}\delta_{\a\b}+\chi^{(2)}_{\a\b},
\ee where $\phi^{(2)}$ is the second order scalar perturbations
and
 $\chi^{(2)}_{\a\b}$ is trace-free. In order to obtain
a simpler form for the second order perturbation equations we
shall not, as in the first order case, separate the trace-free
part of the $\chi^{(2)}_{\a\b}$ perturbations. Substituting
(\ref{mare32}) in (\ref{Ray}) we obtain\cite{MMB,Mena-thesis}
the following evolution equation for $\phi^{(2)}$
\begin{eqnarray}
\label{lp0} &&{\phi_{\scriptscriptstyle}^{(2)}}''+ {a' \over a}
{\phi_{\scriptscriptstyle}^{(2)}}' - \frac{\rho_b a^2}{2}
\phi_{\scriptscriptstyle}^{(2)} = {1 \over 6}
{\gamma^{(1)\a\b}}' \biggl(
  2{a'  \over a}
\gamma^{(1)}_{{\scriptscriptstyle}\a\b}-{\gamma^{(1)}_{{\scriptscriptstyle}\a\b}}' \biggr)
+ \nonumber
\\
&&
+{1 \over 6} \biggl[ 2 \gamma^{(1)\a\b}_{\scriptscriptstyle}
\biggl( 2 \gamma^{(1)\delta}_{{\scriptscriptstyle}\a,\delta\b} 
%\biggr.
%\\
%&&\biggl.
-\nabla^2 \gamma^{(1)}_{{\scriptscriptstyle}\a\b} -
\gamma^{(1)\delta}_{{\scriptscriptstyle}\delta,\a\b} \biggr) -
\gamma^{(1)\delta}_{{\scriptscriptstyle}\delta}
\biggl(\gamma^{(1)\a\b}_{{\scriptscriptstyle},\a\b} - \nabla^2
\gamma^{(1)\a}_{{\scriptscriptstyle}\a} \biggr)
\biggr] +
\\
&&
+\frac{\rho_b a^2}{6} \biggl[{1 \over 4}
\biggl(\gamma^{(1)\a}_{{\scriptscriptstyle}\a} -
\gamma^{(1)\a}_{{\scriptscriptstyle}0\a} \biggr)^2 
%\biggr.\\
%&&\biggl.
+{1 \over 2}
\biggl(\gamma^{(1)\a\b}_{\scriptscriptstyle}
\gamma^{(1)}_{{\scriptscriptstyle}\a\b} -
\gamma^{(1)\a\b}_{{\scriptscriptstyle}0}
\gamma^{(1)}_{{\scriptscriptstyle}0\a\b} \biggr) - \delta_0
\biggl(\gamma^{(1)\a}_{{\scriptscriptstyle}\a} -
\gamma^{(1)\a}_{{\scriptscriptstyle}0\a} \biggr)
\biggr],\nonumber
\end{eqnarray}
where the subscript $"0"$ denotes the evaluation at an initial
time $\tau=\tau_0$.
%%%%%%%%%%%%%%%%%%%%%%%%%%%%%%%%%%%%%%%%%%%%%
\section{Exact Asymptotic Results}
%%%%%%%%%%%%%%%%%%%%%%%%%%%%%%%%%%%%%%%%%%%%%%
It is well known that the first order perturbations of flat dust
FLRW models asymptotically tend to constants in time.\cite{Barrow} 
Now, in order to test the stability of this result
with respect to second order perturbations we can solve the
asymptotic form of (\ref{lp0}). Keeping the lowest order terms on
the right hand side of Eq.~(\ref{lp0}) we have \be \label{2ndphi}
{\phi_{\scriptscriptstyle}^{(2)}}''- {1 \over \tau}
{\phi_{\scriptscriptstyle}^{(2)}}'+\frac{1}{2}\sqrt{\frac{\Lambda}{3}}
\rho_0\tau\phi^{(2)}=F({\bf x})\tau+O(\tau^2), \ee where $F$ is a
spatial function. The solutions to (\ref{2ndphi}) have the form
\be \label{mare38} \phi^{(2)}(\tau,{\bf x})=\frac{F({\bf
x})}{A^2}+ \tau C_3({\bf
x})J\left(\frac{2}{3},\frac{2}{3}A\tau^{3/2}\right)+ \tau C_4({\bf
x})Y\left(\frac{2}{3},\frac{2}{3}A\tau^{3/2}\right), \ee where
$C_3$ and $C_4$ are arbitrary functions and $J$ and $Y$ are Bessel
functions. Therefore, asymptotically\footnote{Note that
$t\to\infty$ corresponds to $\tau\to 0$.} $\phi^{(2)}$ approaches
a constant
value in time.% that depends
%on $\tau_0$, $C_1$, $C_2$, $a_{\a\b}$ and $b_{\a\b}$.

The evolution equation for $\chi^{(2)}_{\a\b}$ can be obtained
from (\ref{evolutiona}) (see\cite{Mena-thesis}) and its
asymptotic form (for each $\a$ and $\b$) is given by \be
\label{mare41} \chi^{(2)''}_{\a\b}-\frac{2}{\tau}
\chi^{(2)'}_{\a\b}+k^2\chi^{(2)}_{\a\b}=
 E_{\a\b}(k)+O(\tau),
\ee where $E_{\a\b}$ depends on $k$. This equation can be solved
to give \be \label{mare42}
\chi^{(2)}_{\a\b}(\tau,k)=\frac{E_{\a\b}(k)}{k}+c_{\a\b}(k\tau\cos(k\tau)-\sin(k\tau))+
d_{\a\b}(k\tau\sin(k\tau)+\cos(k\tau)), \ee where $c_{\a\b}$ and
$d_{\a\b}$ are arbitrary. As a result, as $\tau\to 0$,
$\chi^{(2)}_{\a\b}$ tends to a constant in time which depends on
the asymptotic values of $\phi^{(2)}$ and $d_{\a\b}$. On the other
hand, these values are related to the first order initial free
data
so, at second order there are four independent functions.%, namely $C_1, C_2, a_{\a\b}$ and $b_{\a\b}$,
% which correspond to the first order [FILIPE: OK?] initial data.

To summarise, we showed that second order scalar and
tensor perturbations asymptotically approach a constant value in
time. This demonstrates that the asymptotic behaviour of the first
order perturbations is stable to the presence of second order
perturbations. In addition, using an invariant analysis (see\cite{BMT2001}; cf. also\cite{Starobinsky83}), we
established the validity of the cosmic no-hair conjecture in
nonlinear settings.
%%%%%%%%%%%%%%%%%%%%%%%%%%%%%%%%%%%%%%%%%%%%%
\section{Numerical Results}
\begin{figure}[!htb]
\centerline{\def\epsfsize#1#2{0.5#1}\epsffile{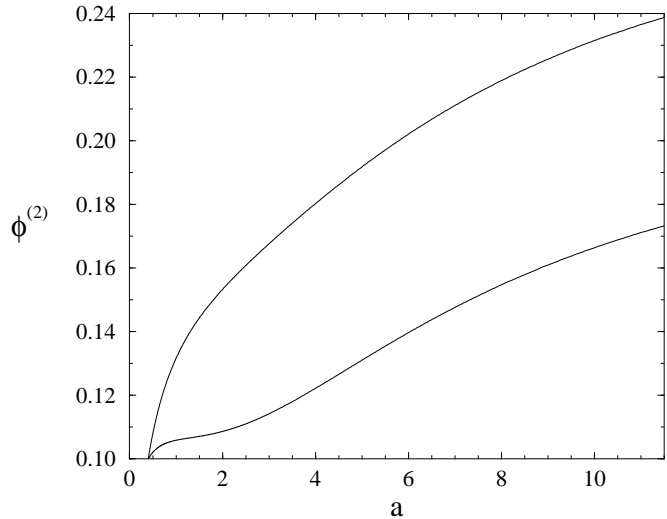}}
\caption{\label{lambda=0.01} Phase portrait representing the
evolution of second order scalar perturbations $\phi^{(2)}$ in a
comoving gauge for a flat FLRW dust model with $\Lambda =0.01$.}
\end{figure}

%%%%%%%%%%%%%%%%%%%%%%%%%%%%%%%%%%%%%%%%%%%%%
Our analysis so far has concentrated on the asymptotic behaviour
of the perturbations. The all time evolution of the second order
perturbations is of potential importance for processes that evolve
in time, such as the structure formation process. Here as a first
step in this direction we numerically integrated the equation
(\ref{lp0}) for several initial conditions  in order to
investigate the evolution of $\phi^{(2)}$ in time. We found that
even though asymptotically there is a single future attractor that
determines the qualitative behaviour of $\phi^{(2)}$ at late
times, the precise approach to this attractor can be different at
intermediate times, as shown in Figure \ref{lambda=0.01}. The two
curves represent the evolution of $\phi^{(2)}$ for different
initial conditions. Both curves have initial values
$\Lambda=0.01$, $a(\tau_0)=0.9$, $\phi^{(2)}(\tau_0)=0.1$,
$\phi^{(1)}(\tau_0)=0.01$, $\phi^{(1)'}(\tau_0)=0.01$ and
$\rho_0=0.01$. The lower curve has $\phi^{(2)'}(\tau_0)=0.01$, and
the upper curve $\phi^{'(2)}(\tau_0) = 0.025$. Varying
$\phi^{'(2)}(\tau_0)$ between $0.1$ and $0.025$ results in a
family of curves lying between the two depicted curves. As can be
seen, the evolution of $\phi^{(2)}$ can have distinct behaviours
over intermediate times; having for example either one or three
inflection points depending on the initial data.
%____________________________________________________________________________
%____________________________________________________________________

%Figure \ref{lambda=0.01} indicates that after
%the initial phase, where $\rho$ dominates the
%dynamics of the evolution equation, there is an intermediate phase in the evolution
%of $\phi^{(2)}$ where the cosmological constant dynamically competes with $\rho$ forcing,
%in some cases, the existence
%of an almost flat region represented in the lower curve of Figure \ref{lambda=0.01}. The late
%time dynamics is dominated by $\Lambda$ and will be studied below.
This is an example of a dynamical system with the same
attractor but different transient phases.% and it would be
%interesting in future to study it quantitatively.
 This type of behaviour could result in different rates of
structure formation over intermediate time scales, depending upon
the choice of the initial data. In particular we could have a
possible intermediate period where $\phi^{(2)}$ can be almost
constant in time, as is the case with the lower curve in Figure
\ref{lambda=0.01}.
%%%%%%%%%%%%%%%%%%%%%%%%%%%%%%%%%%%%%%%%%%%%%%
\section{Discussion and Conclusions}
%%%%%%%%%%%%%%%%%%%%%%%%%%%%%%%%%%%%%%%%%%%%%%
We have summarised our very recent results\cite{BMT2001,Mena-thesis} 
on the evolution of second order
perturbations in flat irrotational dust FLRW models. We have shown
that second order perturbations tend asymptotically to constants
in agreement with the cosmic censorship conjecture.

In order to obtain the behaviour of such perturbations at
intermediate times, we have also integrated numerically the scalar
perturbation equation and found different transient behaviours
over intermediate time scales, depending upon the choice of the
initial conditions. This can have interesting consequences for
nonlinear phenomena that evolve in time, such as structure
formation.
%%%%%%%%%%%%%%%%%%%%%%%%%%%%%%%%%%%%%%%%%%%%%%%
\section*{Acknowledgements}
%%%%%%%%%%%%%%%%%%%%%%%%%%%%%%%%%%%%%%%%%%%%%%%
FCM thanks CMAT (Universidade do Minho) and the Organizing
Committee of the "5th Friedmann seminar" for support. RT thanks
CBPF for hospitality.
%Rio de Janeiro.

%%%%%%%%%%%%%%%%%%%%%%%%%%%%%%%%%%%%%%%%%%%%%%%%%%

\end{document}